\documentclass[pss,fleqn]{w-art}
\usepackage{times}
\usepackage{w-thm}
\usepackage{w-greek}
\usepackage{amsmath}
\usepackage{amssymb}
\usepackage{cite}
\theoremstyle{plain}

\theoremstyle{definition}

\usepackage[]{graphicx}
\chardef\bslash=`\\ 

\begin{document}
\DOIsuffix{theDOIsuffix}
\Volume{XX}
\Issue{1}
\Month{01}
\Year{2003}
\Receiveddate{15 March 2004}
\Reviseddate{26 April 2004}
\Accepteddate{26 April 2004}
\Dateposted{25 June 2004}

\title[Critical properties]{Critical properties in long-range
hopping Hamiltonians}

\author[Sh. First Author]{E. Cuevas\footnote{Corresponding
     author: e-mail: {\sf ecr@um.es}}
     \inst{}} \address[\inst{}]{Departamento de F{\'\i}sica,
     Universidad de Murcia, E-30071 Murcia, Spain.}

\begin{abstract}
Some properties of $d$-dimensional disordered models with long-range
random hopping amplitudes are investigated numerically at criticality.
We concentrate on the correlation dimension $d_2$ (for $d=2$) and the
nearest level spacing distribution $P_c(s)$ (for $d=3$) in both the weak
($b^d \gg 1$) and the strong ($b^d \ll 1$) coupling regime, where the
parameter $b^{-d}$ plays the role of the coupling constant of the model.
It is found that (i) the extrapolated values of $d_2$ are of the form
$d_2=c_db^d$  in the strong coupling limit and $d_2=d-a_d/b^d$ in the
case of weak coupling, and (ii) $P_c(s)$ has the asymptotic form
$P_c(s)\sim\exp (-A_ds^{\alpha})$ for $s\gg 1$, with the critical exponent
$\alpha=2-a_d/b^d$ for $b^d \gg 1$ and $\alpha=1+c_d b^d$ for $b^d \ll 1$.
In these cases the numerical coefficients $A_d$, $a_d$ and $c_d$ depend only
on the dimensionality.    
\end{abstract}
\maketitle

\section{Introduction}
\label{int}

Quantum phase transitions in disordered electronic systems remain one of
the central problems in conden\-sed-matter physics. Considerable attention
has recently focused on both the energy levels and the critical eigenfunctions,
which strongly fluctuate near the critical point and thus have multifractal
scaling properties \cite{HP83,HJ86,Ja94,FE95,Mi00,EM00,CO02}. Wave-function
statistics can be characterized through the set of generalized fractal
dimensions $d_q$ which are associated with the scaling of the $q$-th moment
of the wave-function intensity. A complete knowledge of $d_q$ is equivalent
to a complete physical characterization of the fractal \cite{HP83}.

Among all the fractal dimensions, the correlation dimension $d_2$ of the
eigenfunctions plays the most prominent role, since it has been related
to the level compressibility \cite{CK96,KM97,Kr99,KT00}, the spatial
dispersion of the diffusion coefficient \cite{Ch88} and the anomalous
spreading of a wave-packet at the mobility edge \cite{HS94}.
Multifractality is also related to the probability overlap of eigenstates
with energy separation much greater than the average separation level. In
particular, $d_2$ describes these density correlations. It has been shown
that for multifractal eigenstates these correlations decay slowly no matter
how sparse those states are \cite{FM97}.

Short-range correlations in the energy levels, which are closely related
to the localization properties of the corresponding wave-functions
\cite{AS86,SS93,KL94}, can be described by the critical distribution
$P_c(s)$ of the normalized spacings $s$. The large $s$ asymptotic behavior
of this distribution is still an open question and the subject of
discussion \cite{Cu03b}.

Metal-insulator transitions (MIT's) depend on the dimensionality and
symmetries of the system and can occur in both the strong disorder and
the weak disorder regime (strong-coupling or weak-coupling regime,
respectively, of the corresponding field-theoretical description) as
well as in the intermediate regime. Each regime is characterized by its
respective coupling strength \cite{Ef83}.

The disorder-induced MIT is usually investigated for Hamiltonians with
short-range, off-diagonal matrix elements (e.g., the canonical Anderson
model). Other Hamiltonians exhibiting an MIT in arbitrary dimension $d$
are those that include long-range hopping terms \cite{PS98,PS02,An58,Le89,
Le99,AL97}. The effect of long-range hopping on localization was originally
considered by Anderson \cite{An58} for randomly distributed impurities in
$d$ dimensions with the $V(\boldsymbol{r}-\boldsymbol{r'}) \sim |\boldsymbol{r}-
\boldsymbol{r'}|^{-\beta}$ hopping interaction. It is known \cite{An58,Le89,AL97}
that all states are extended for $\beta<d$, whereas for $\beta > d$ the states
are localized. Thus, the MIT can be studied by varying the exponent $\beta$ at
fixed disorder strength. At the transition line $\beta=d$, a real-space
renormalization group can be constructed for the distribution of couplings
\cite{Le89,Le99}. These models are the most convenient for studying critical
properties numerically since the exact quantum critical point is known
($\beta=d$) and, in addition, they allow the low-dimensional cases to be treated,
thus using larger system sizes and reducing the numerical effort.

Although a great progress in understanding critical properties of the
$1d$ long-range random hopping Hamiltonian has recently been made
\cite{Mi00,CO02,KM97,KT00,MF96,ME00,VB00,Va02,CG01,Cu02,YK03,YK03a,Ga03,
Cu03,Cu03a,PS94} explicit results for the $2d$ and $3d$ systems are still
lacking. Our aim was to investigate the two previously mentioned important
quantities, the correlation dimension and nearest level spacing distribution,
at the MIT, which have been left unexplored. The last quantity could help us
to solve the existing controversy about the form of $P_c(s)$ at the MIT.

The paper begins by first giving the model used for the calculations in
Sec. \ref{mod}. The results for the correlation dimension and for the
critical level spacing distribution are presented in Sec. \ref{crd} and
\ref{lsd}, respectively. Finally, Sec. \ref{sum} summarizes our findings.

\section{The model}
\label{mod}

In order to fully represent the mesoscopic systems we introduce an explicit
dependence on dimensionality $d$ in the widely studied power-law random banded
matrix (PRBM) ensemble \cite{Mi00,CO02,KM97,KT00,MF96,ME00,VB00,Va02,CG01,Cu02,
YK03,YK03a,Ga03,Cu03,Cu03a} (for closely related models see also Ref. \cite{PS94}).
Thus, we consider a generalization to $d$ dimensions of this ensemble. The
corresponding Hamiltonian, which describes non-interacting electrons on a
disordered $d$-dimensional square lattice with random long-range hopping, is
represented by real symmetric matrices, whose entries are randomly drawn from
a normal distribution with zero mean, $\left\langle {\cal H}_{ij} \right\rangle =0$,
and a variance which depends on the distance between the lattice sites
$\boldsymbol{r}_i$
\begin{equation}
\left\langle |{\cal H}_{ij}|^2\right\rangle =\frac{1}
                     {1+(|\boldsymbol{r}_i-\boldsymbol{r}_j|/b)^{2\beta}}
                    \left\{\begin{array}{ll}
                    \dfrac{1}{2}     \ ,\quad & i\neq j\\
                    1 \ ,\quad & i=j
       \end{array}\right.
\label{h1dor}\;.
\end{equation}

Using field theoretical methods \cite{Mi00,EM00,KM97,KT00,Kr99,Le89,AL97,MF96},
the PRBM model was shown to undergo a sharp transition at $\beta=d$ from
localized states for $\beta>d$ to delocalized states for $\beta<d$. This
transition shows all the key features of the Anderson MIT, such as multifractality
of the eigenfunctions and non-trivial spectral compressibility at
criticality. In what follows, we focus on the critical value $\beta=d$.
The parameter $b^d$ in Eq. (\ref{h1dor}) is an effective bandwidth that
serves as a continuous control parameter over a whole line of criticality,
i.e, for an exponent equal to $d$ in the hopping elements
${\cal H}_{ij} \sim b^d$ \cite{Le89}. Furthermore, it determines the critical
dimensionless conductance in the same way as the dimensionality labels the
different Anderson transitions. Each regime is characterized by its respective
coupling strength, which depends on the ratio
$(\langle |{\cal H}_{ii}|^2 \rangle/\langle |{\cal H}_{ij}|^2 \rangle)^{1/2}
\propto b^{-d}$ between diagonal disorder and the off-diagonal transition
matrix elements of the Hamiltonian \cite{Ef83}.

Many real systems of interest can be described by Hamiltonians (\ref{h1dor}).
Among such systems are optical phonons in disordered dielectric materials
coupled by electric dipole forces \cite{Yu89}, excitations in two-level systems
in glasses interacting via elastic strain \cite{BL82}, magnetic impurities in
metals coupled by an $r^{-3}$ Ruderman-Kittel-Kasuya-Yodida interaction
\cite{CB93}, and impurity quasiparticle states in two-dimensional disordered
$d$-wave superconductors \cite{BS96}. It also describes a particle moving fast
through a lattice of Coulomb scatterers with power-law singularity \cite{AL97},
the dynamics of two interacting particles in a $1d$ random potential \cite{PS97}
and a quantum chaotic billiard with a non-analytic boundary \cite{CP99}.

The two limiting cases of the $1d$ model, $b \gg 1$ and $b \ll 1$, which
correspond to the weak and the strong disorder limits, respectively, can
be studied via the mapping onto the supermatrix $\sigma$-model for $b$ large
\cite{Mi00,MF96} and using the renormalization-group method of Refs.
\cite{Le89,Le99} for small $b$. In particular, one finds the following result
for the correlation dimension $d_2$ at the center of the band

\begin{equation}
d_2=\left\{\begin{array}{ll}
                     1-\dfrac{1}{\pi b} \ ,\quad & b \gg 1\\
                     2 b \; ,\quad & b \ll 1
       \end{array}\right.
\label{d2vb}\;.
\end{equation}
Thus, for the $1d$ version of Hamiltonian (\ref{h1dor}), which possesses
the line of critical points $0 < b < \infty$, $d_2$ changes from $0$ to
the system dimensionality $d=1$ as $b$ increases. Eq. (\ref{d2vb}) has
been numerically confirmed by several groups using exact diagonalization
\cite{EM00,ME00,Va02,Cu03a}. We stress that, unlike the $1d$ PRBM model,
it has not until now been possible to analytically solve the $2d$ and $3d$
disordered models with long-range transfer terms.

The system sizes used are $L=24$, $36$, $48$, $72$ and $96$ in $2d$,
and $L=8$, $12$, $14$ and $16$ in $3d$, whereas  $b^d$ ranges in the
interval $0.01 \le b^d \le 10$. We consider a small energy window, containing
about 10\% of the states around the center of the spectral band. The number
of random realizations is such that the number of critical levels and
eigenstates included for each $L$ is roughly $1.3\times 10^6$, except for
the larger system size in both dimensions, for which this number is about
$6\times 10^5$. In order to reduce edge effects, periodic boundary
conditions are included. The power-law nature of Hamiltonians (\ref{h1dor})
did not allow us to use efficient algorithms, such as the Lanzos algorithm,
which is usually applied to the study of the MIT in the Anderson model, due
to the large degree of sparcity of the corresponding Hamiltonian. Instead,
we use standard diagonalization subroutines.

\section{The correlation dimension}
\label{crd}

In this section we investigate critical fluctuations in the eigenfunctions
of the two-dimensional model (\ref{h1dor}) in terms of multifractal measures,
focusing on the correlation dimension $d_2$. At the MIT, where the natural
length scale (the localization length) diverges, strong fluctuations in the
wave-functions $\psi_\mu(\boldsymbol{r})$ appear on all length scales. These
fluctuations can be characterized by a set of inverse participation ratios (IPR)
\cite{We80}
\begin{equation}
I_{\mu}(q)=\int_\Omega d^2r\;|\psi_\mu(\boldsymbol{r})|^{2q}\
\propto L^{-\tau_q}\;, \quad \tau_q=d_q(q-1) \;,\label{ipr}
\end{equation}
where $d_q$ is a set of generalized fractal dimensions. The index $\mu$
labels different eigenfunctions and $\Omega$ denotes a $2$-dimensional
region with linear dimension $L$. Equation (\ref{ipr}) is valid for
individual states and for their ensemble average since the spectrum of
multifractal dimensions has universal features for states in the vicinity
of the MIT \cite{Ja94}. The inverse of $I_{\mu}(2)$ roughly equals the
number of nonzero wave-function components, for which reason it is a widely
accepted measure of the extension of the states. Note that in a good metal,
for which eigenfunctions are ergodic, the IPR scale with size $L$ as
$I_{\mu}(q) \propto L^{-d(q-1)}$, whereas in an insulator, with localized
states, $I_{\mu}(q) \propto L^{0}$.

For the computation of $\tau_q$ we used the standard box-counting procedure
\cite{Ja94}, first dividing the system of $L^2$ sites into $N_l=(L/l)^2$
boxes of linear size $l$ and determining the box probability of the wave
function in the $i$ box by
$p_i(l)=\sum_{\boldsymbol{r}} |\psi_\mu(\boldsymbol{r})|^2$, where the
summation is restricted to sites within that box, and $\psi_\mu(\boldsymbol{r})$
denotes the amplitude of an eigenstate with energy $\epsilon_\mu$ at site
$\boldsymbol{r}$. The normalized $q$-th moments of this probability constitute a
measure. From this, the mass exponents $\tau_q(L)$, which encode generalized
dimensions $d_q(L)=\tau_q(L)/(q-1)$, can be obtained \cite{CJ89}
\begin{equation}
\tau_q(L)=\lim_{\delta \to 0} \frac {\ln \displaystyle \sum_{i=1}^{N_l}
p_i^q(l)}{\ln \delta}\;,\label{tauq}
\end{equation}
where $\delta=l/L$ denotes the ratio of the box sizes and the system size.
It should be made clear that the calculation of $\tau_q(L)$ is suitable only
if the conditions \cite{Ja94}
\begin{equation}
a\ll l < L \ll \xi\ \;, \label{ineq}
\end{equation}
are satisfied, where $\xi$ is the localization or correlation length
and $a$ is the lattice spacing (or any microscopic length scale of
the system). In practice, $\tau_q(L)$ is found by performing a linear
regression of $\ln \sum_{i=1}^{N_l}p_i^q(l)$ with $\ln \delta$ in a finite
interval of $\delta$. In order to properly satisfy the previous conditions
(\ref{tauq}) and (\ref{ineq}), we take $\delta$ to be in the interval
$(0.1, 0.4)$. Since we are mainly interested in the correlation dimension,
we shall restrict ourselves to the value $q=2$ and so $d_2=\tau_2$.

\begin{figure}[t]
\begin{minipage}[t]{0.48\textwidth}
\includegraphics[width=\textwidth]{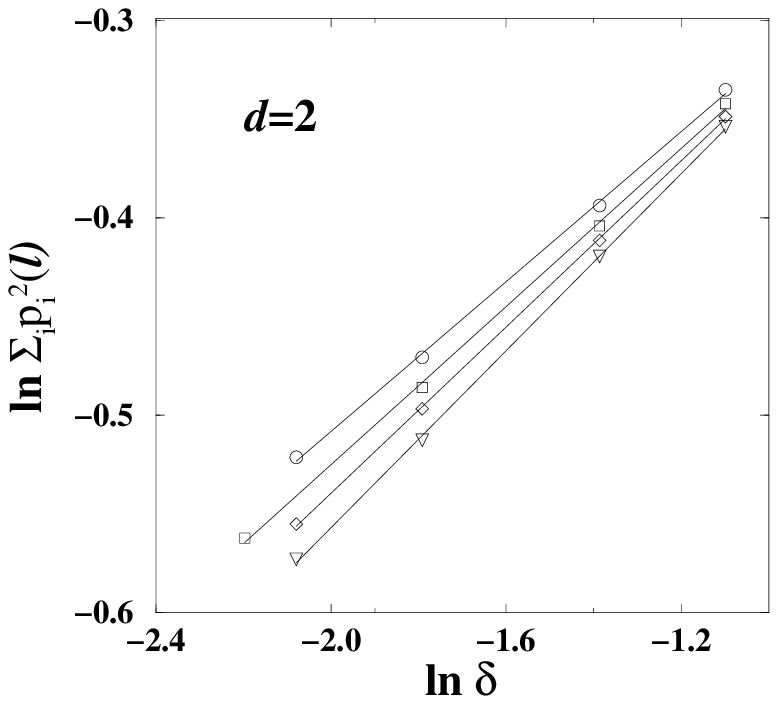}
\caption{$\ln \delta$ dependence of $\ln \displaystyle\sum_{i=1}^{N_l}p_i^2(l)$
in the strong-coupling regime ($b^2=0.05$) for different system sizes: $L=24$
(circles), $36$ (squares), $48$ (diamonds) and $96$ (triangles). The straight
lines whose slopes correspond to the values of $d_2(L)$ are linear fits to
Eq. (\ref{tauq}).}
\label{fig_1}
\end{minipage}
\hfil
\begin{minipage}[t]{0.48\textwidth}
\includegraphics[width=\textwidth]{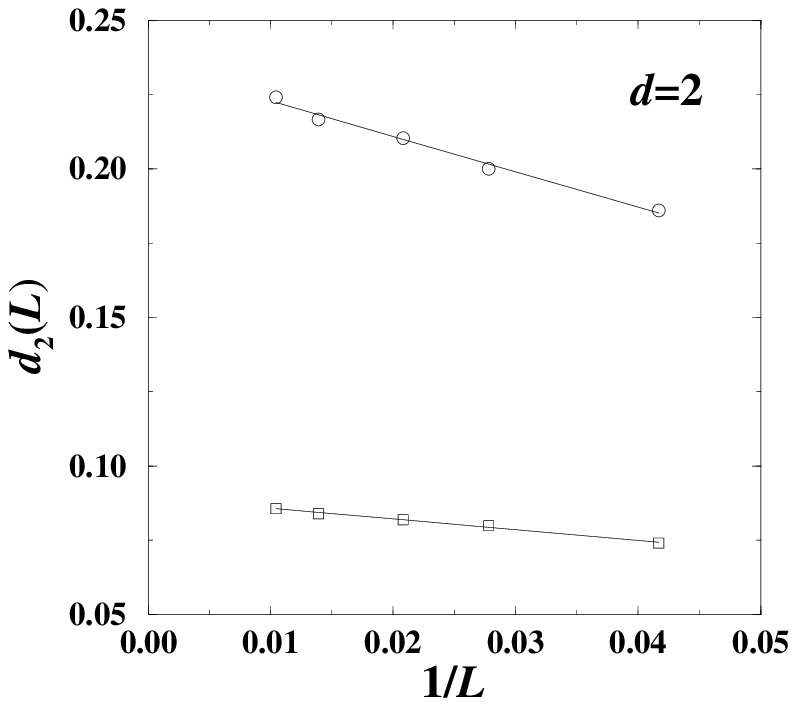}
\caption{Finite-size corrections to the correlation dimension $d_2(L)$ in
the strong-coupling regime $b^2=0.02$ (squares) and $b^2=0.05$ (circles);
solid lines are fits to Eq. (\ref{s2l}).}
\label{fig_2}
\end{minipage}
\end{figure}

Using the exact eigenstates of Hamiltonian (\ref{h1dor}) obtained from
numerical diagonalizations, we evaluate, for each value of $b$ and $L$,
the numerator on the right-hand side of Eq. (\ref{tauq}) for decreasing
box sizes, and then calculate $d_2(L)$ from the slope of the graph of the
numerator vs $\ln \delta$. Figure 1 provides an example of the $\ln \delta$
dependence of $\ln \displaystyle\sum_{i=1}^{N_l}p_i^2(l)$ in the
strong-coupling regime ($b^2=0.05$) for different system sizes:
$L=24$ (circles), $36$ (squares), $48$ (diamonds) and $96$ (triangles).
Clearly, there is no ambiguity in the determination of the slopes that
correspond to the values of $d_2(L)$. These slopes are summarized in Fig. 2.

In Ref. \cite{Cu03a} it was shown that the finite-size corrections to $\tau_2(L)$
are of power-law type ($\sim L^{-1}$) for the $1d$ PRBM model. A similar decay
was found for the multifractal spectrum $f(\alpha)$ and its singularity strength
$\alpha$ in $d=1$ and $2$ \cite{Cu03}. Thus, in order to predict the asymptotic
values of $d_2$, a curve of the form
\begin{equation}
d_2(L)=d_2+a_2/L \;, \label{s2l}
\end{equation}
is proposed. Other forms, such as exponential ($\sim e^{-a_2 L}$) or inverse
logarithmic ($\sim 1/\ln a_2 L$) have been rejected since we have checked that
none of them can adequately describe the $d_2$ size behavior.

In Fig. 2, we represent the finite-size corrections for the correlation
dimension $d_2(L)$ of the $2d$ disordered system described by Eq. (\ref{h1dor})
in the strong-coupling regime $b^2=0.02$ (squares) and $0.05$ (circles). Note that
significant finite-size effects are present. The straight lines are linear fits to
Eq. (\ref{s2l}) and intercept the vertical axis at $0.0894 \pm 0.0004$ and
$0.235 \pm 0.002$, respectively. 

\begin{figure}[t]
\begin{minipage}[t]{0.48\textwidth}
\includegraphics[width=\textwidth]{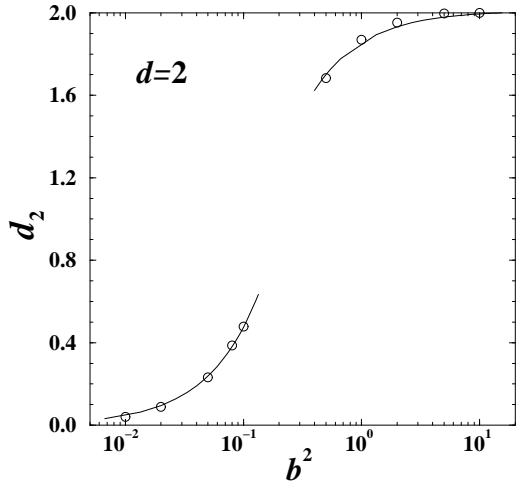}
\end{minipage}
\hfil
\begin{minipage}[t]{0.48\textwidth}
\vskip -6.8cm
\caption{$b^2$ dependence of the correlation dimension $d_2$ (circles)
for the $2d$ disordered model. Solid lines are fits to Eq. (\ref{d2vb2})
corresponding to the limiting cases of weak ($b^2 \gg 1$) and strong
($b^2 \ll 1$) disorder.}
\label{fig_3}
\end{minipage}
\end{figure}

The extrapolated values of $d_2$, as obtained from the previous fits, are shown
in Fig. \ref{fig_3} (circles) as a function of the disorder parameter $b^2$ of
the PRBM model. They clearly change continuously from 0 as $b^2 \to 0$ to the
system dimensionality $d=2$ as $b^2 \to \infty$. In the two limiting cases of
weak ($b^2 \gg 1$) and strong ($b^2 \ll 1$) disorder regimes $d_2$ can be well
fitted by
\begin{equation}
d_2=\left\{\begin{array}{ll}
                     2-\dfrac{a_2}{b^2} \ ,\quad & b^2 \gg 1\\
                     c_2 b^2 \; ,\quad & b^2 \ll 1
       \end{array}\right.
\label{d2vb2}\;,
\end{equation}
respectively. These fits are shown as solid lines in Fig. \ref{fig_3}. The
fitting parameters are $a_2=0.16 \pm 0.06$ and $c_2=4.76 \pm 0.07$. Note
the similarity of Eqs. (\ref{d2vb2}) with the corresponding to the $1d$
case,  Eqs. (\ref{d2vb}).

\section{The level spacing distribution}
\label{lsd}

This section is devoted to the analysis of short-range energy level
correlations in the three-dimensional model. The emphasis is on the large $s$
asymptotic behavior of the level spacing distribution at the mobility edge.
We first review existing analytical results concerning this distribution.

On the localized side of the transition, states with close energy levels
are typically localized at different parts of space and have an exponentially
small overlap. Their levels are therefore uncorrelated and the corresponding
spacings are distributed according to the Poisson law
\begin{equation}
P_{\rm P}(s) = \exp(-s)\;.\label{poisson}
\end{equation}

In contrast, in the metallic regime, the large overlap of
delocalized states, which are essentially structureless, induces
correlations in the spectrum, leading to the well known level
repulsion effect. If the system is invariant under rotation and
under time-reversal symmetry (orthogonal symmetry), the normalized
spacings $s$ follow Wigner-Dyson statistics at the infinite system
size limit
\begin{equation}
P_{\rm W}(s) = \frac{\pi}{2}s\exp \left (-\frac{\pi}{4}s^2 \right )\;.
\label{wigner}
\end{equation}

At the disorder-induced MIT, the statistics of energy levels changes
drastically and presents distinct features reflecting criticality of
the theory. This statistics is characterized by a third universal
(i.e., independent of the system size and of the details of the
Hamiltonian model) distribution $P_c(s)$, which is different from both
Wigner-Dyson statistics and the Poisson statistics \cite{SS93,AZ88}.
On the one hand, the influence of the MIT on the spectral properties
was studied in Refs. \cite{SS93,AZ88} by means of the impurity diagram
technique combined with scaling assumptions. In these studies, it was
conjectured that
\begin{equation}
P_c(s)\sim \exp (-\kappa s)\;, \quad s \gg 1 \;,
\label{tailsh}
\end{equation}
with $\kappa \approx 3.3$, the reason for such behavior being that the
Thouless energy at the transition point is of the order of the average
level spacing ($\epsilon_c/\Delta \approx 1$), and so the levels's
repulsion is effective only for $s \lesssim 1$.

A short-range plasma model with interaction only between closest
neighbors \cite{GJ98,JK99,BG99,AJ01} suggest that the universality
connected with the spectral fluctuations at the MIT is the intermediate
spectral statistics $P_c(s)=4s\exp (-2 s)$.

On the other hand, by mapping the energy level distribution onto the
Gibbs distribution for a classical one-dimensional gas with a repulsive
pairwise interaction, Ref. \cite{AK94} derived the following asymptotic
form for $P_c(s)$
\begin{equation}
P_c(s)\sim \exp (-A_ds^{\alpha})\;, \quad s \gg 1\;,
\label{tailar}
\end{equation}
where the coefficient $A_d$ depends only on the dimensionality $d$,
and where the critical exponent $\alpha$, which ranges in the interval
$1<\alpha<2$, is related to the correlation length exponent $\nu$
and to the dimensionality through $\alpha=1+(d\nu)^{-1}$.

Neither as regards the numerical description of $P_c(s)$ is there any
consensus. The exponential decay, Eq. (\ref{tailsh}), of $P_c(s)$ has
been confirmed by most groups at different MIT's (see Refs.
\cite{Ev95,Ho96,KO96,ZB97,Ni99} and references therein), while an
exponent $\alpha \approx 1.2$ has been found in Refs.
\cite{Ev94,VH95} from a fit in the whole range of spacings to a
distribution of the form $P_c(s)=Bs\exp(-As^{\alpha})$ or, indirectly,
from the two-point correlation function of the density of states \cite{BM95}.
Anyway, the behavior (\ref{tailar}) with some nontrivial $1 \le \alpha \le 2$
is what one would expect at the mobility edge.

It should be pointed out that MIT's generically take place at strong disorder
(conventional Anderson transition, quantum Hall transition, transition
in $d=2$ for electrons with strong spin-orbit coupling, etc.). In this
regime, the predicted \cite{AK94} exponent $\alpha=1+(d\nu)^{-1}$ slightly
deviates from unity, making it relatively difficult to see on the
numerically calculated tails of $P_c(s)$ (e.g, at the standard Anderson
transition in $3d$ $\alpha \approx 1.2$). To overcome this problem, it is
necessary to investigate transitions which occur at the opposite limit
(weak coupling regime). This area has recently been investigated for
the model (\ref{h1dor}) when $d=1$ and $2$ \cite{Cu03b}, and an exponent
$\alpha$ close to the Wigner-Dyson value $\alpha=2$ has been found at large
values of $b$ and $b^2$, respectively.

Here, we make a similar study for the more realistic and interesting case $d=3$.
From results of detailed high precision numerical investigations, we will show
unambiguously that Eq. (\ref{tailar}) is indeed  correct while the validity of
Eq. (\ref{tailsh}) is limited to the case of very strong disorder (strictly at
the limit of infinity coupling strength). In addition, we find that the exponent
$\alpha$ in Eq. (\ref{tailar}) continuously varies between 1 and 2 as the coupling
strength of the Hamiltonian model changes from 0 to $\infty$.

For the computation of $P_c(s)$, we unfold the spectrum in each case to
a constant density, and rescale it so as to have the mean spacing equal
to unity. In order to diminish the magnitude of the relative fluctuations
and to analyze the asymptotic behavior in detail, it is more convenient to
consider the cumulative level spacing distribution function
$I(s)=\int_{s}^{\infty} P(s')ds'$. Note that the integration does not change
the asymptotic behavior of $P(s)$. The Wigner surmise, Eq. (\ref{wigner}),
and the Poisson distribution, Eq. (\ref{poisson}), yield
$I_{\rm W} (s)=\exp (-\pi s^2/4)$ and $I_{\rm P}(s)=\exp (-s)$, respectively.

\begin{figure}[t]
\begin{minipage}[t]{0.48\textwidth}
\includegraphics[width=\textwidth]{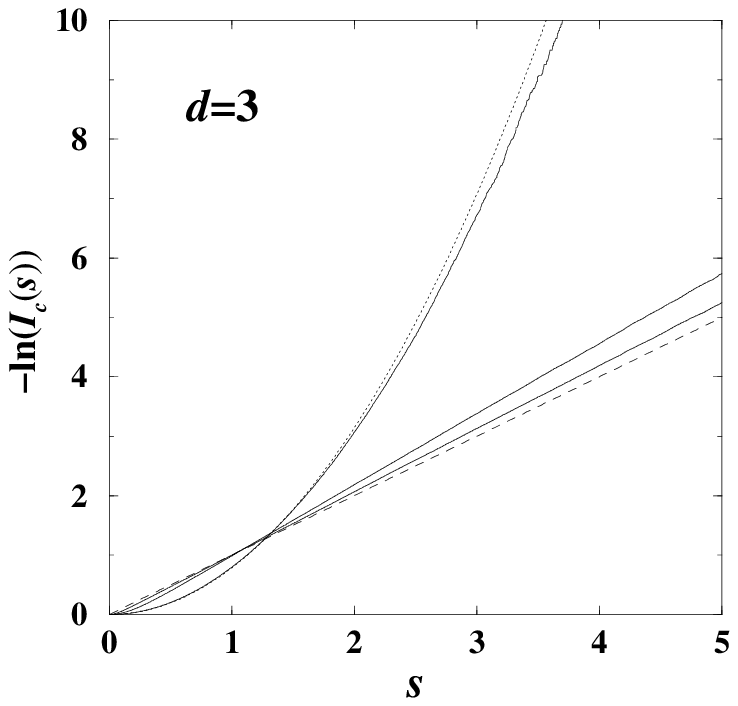}
\caption{Integrated probability $I_c(s)$ of the $3d$ system for $L=14$
at $b^3=0.02$, 0.05 and 5 (from bottom to top). Dotted and dashed lines are
$I_{\rm W}(s)$ and $I_{\rm P}(s)$, respectively.}
\label{fig_4}
\end{minipage}
\hfil
\begin{minipage}[t]{0.48\textwidth}
\includegraphics[width=\textwidth]{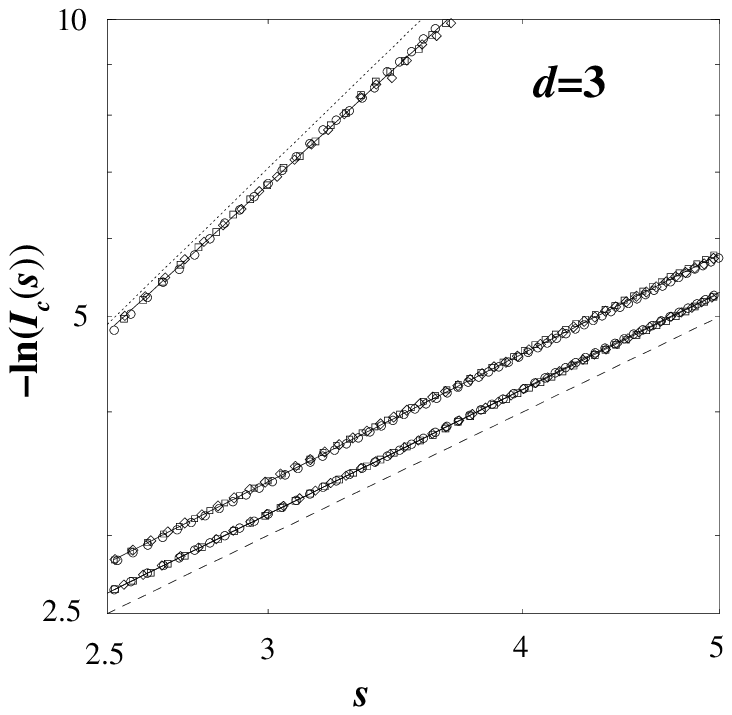}
\caption{Log-log plot of the integrated probability $I_c(s)$ of the $3d$ system
at $b^3=0.02$, 0.05 and 5 (from bottom to top) and different system sizes $L=8$
(circles), $12$ (squares) and $16$ (diamonds). Dotted and dashed lines are
$I_{\rm W}(s)$ and $I_{\rm P}(s)$, respectively, and the straight lines are
fits to Eq. (\ref{tailar}).}
\label{fig_5}
\end{minipage}
\end{figure}

Figure \ref{fig_4} displays our results for the integrated probability
$I_c(s)$ of the $3d$ model for $L=14$ at $b^3=0.02$, 0.05 and 5, which
are depicted consecutively from bottom to top. Dotted and dashed lines,
which correspond to $I_{\rm W}(s)$ and $I_{\rm P}(s)$, respectively, are
given for comparison. A gradual crossover in the large $s$ tail of
$I_{\rm c}(s)$ from the Poisson to the Wigner-Dyson limiting forms as one
increases the inverse coupling constant $b^3$ of the model can clearly be
seen. So, we can therefore expect an exponent $\alpha$
in Eq. (\ref{tailar}), which spans the interval $[1,2]$, in agreement
with Ref. \cite{AK94}. For the $1d$ and $2d$ cases the behavior is quite
similar \cite{Cu03b}. 

Next we consider the behavior of $I_c (s)$ with system size $L$.
The results for $s$ large of the critical $I_c (s)$ for the $3d$ model
at different values of $b^3$ are shown in a log-log scale in
Fig. \ref{fig_5} for different system sizes: $L=8$ (circles), $12$
(squares) and $16$ (diamonds). Note that $I_c(s)$ is an $L$-independent
universal scale-invariant function that interpolates, as previously
mentioned, between Wigner and Poisson limits. This result confirms the
existence of a critical distribution exactly at the transition. Dotted
and dashed lines correspond to $I_{\rm W}(s)$ and $I_{\rm P}(s)$,
respectively. We checked that the normalized variances of $P_c(s)$ are
indeed scale-invariant at each critical point studied \cite{Cu99,CL96}.
The straight line behavior of the data in such
a plot at all values of $b^3$ considered is undoubtedly consistent
with a $b^3$ dependent exponent $\alpha$ in Eq. (\ref{tailar}).
The values of $b^3$ reported are $0.02$, $0.05$ and $5$,
from bottom to top. The best fit to Eq. (\ref{tailar}) in the interval
$2.5 \lesssim s \lesssim 5$ for small $b^3$ and $2.5 \lesssim s \lesssim 4$
for large $b^3$, yields $\alpha=1.008$, 1.039 and 1.901, respectively,
thus confirming the result of \cite{AK94}. Note that for the large energy
ranges considered, where $I_c(s)$ vary by one to three orders of magnitude,
the quality of the fits, which are represented as solid straight lines, is
evident.

\begin{figure}[t]
\begin{minipage}[t]{0.48\textwidth}
\includegraphics[width=\textwidth]{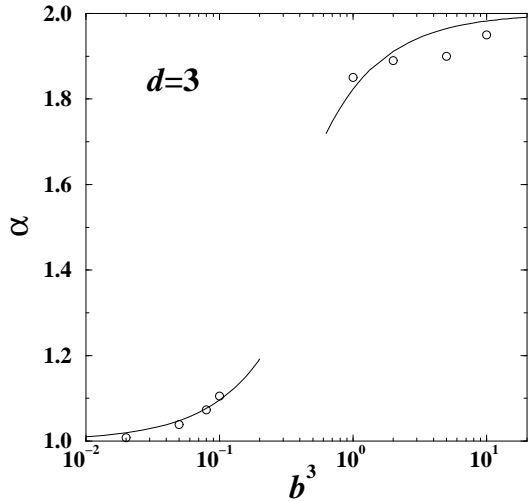}
\end{minipage}
\hfil
\begin{minipage}[t]{0.48\textwidth}
\vskip -7cm
\caption{$b^3$ dependence of the critical exponent $\alpha$ (circles)
for the $3d$ disordered system. Solid lines are fits to Eq. (\ref{alfavbd})
corresponding to the limiting cases of weak ($b^3 \gg 1$) and strong
($b^3 \ll 1$) disorder.}
\label{fig_6}
\end{minipage}
\end{figure}

Finally, the disorder dependence of the critical exponent $\alpha$, as obtained
from the previous fits for the $3d$ system (circles) is shown in Fig. \ref{fig_6}
in the broad range of the parameter $b^3$ of the $3d$ PRBM model. It clearly changes
continuously from the Poisson value $\alpha=1$ as $b^3 \to 0$ to the Wigner-Dyson
value $\alpha=2$ as $b^3 \to \infty$. In the two limiting cases of weak ($b^3 \gg 1$)
and strong ($b^3 \ll 1$) disorder regimes it can be fitted by
\begin{equation}
\alpha=\left\{\begin{array}{ll}
                     2-\dfrac{a_3}{b^3} \ ,\quad & b^3 \gg 1\\
                     1+c_3 b^3 \; ,\quad & b^3 \ll 1
       \end{array}\right.
\label{alfavbd}\;,
\end{equation}
respectively. These fits are shown as solid lines in Fig. \ref{fig_6}.
The fitting parameters are $a_3=0.18 \pm 0.04$ and $c_3=0.96 \pm 0.07$.
Note that at $b^3 \gg 1$ the condition
$|\boldsymbol{r}_i-\boldsymbol{r}_j|/b \gg 1$ is not completely fulfilled
for the system sizes considered and for the largest $b^3$ reported the
corresponding $\alpha$ saturate at smaller values than predicted by the
previous equation. From eq. (\ref{alfavbd}), the Poissonian tail of $P_c(s)$,
eq. (\ref{tailsh}), is recovered for large spacings at the limit of very strong
coupling $b^3 \to 0$. So, we conclude that in the case of very strongly coupled
Hamiltonians only, eq. (\ref{tailar}) loses its validity  and eq. (\ref{tailsh})
applies.

\section{Summary}
\label{sum}

We have calculated the correlation dimension $d_2$ of the eigenfunctions
and the nearest level spacing distribution $P_c(s)$ of non-interacting
electrons on $d$-dimensional disordered models ($d=2$ and $3$) with long-range
random transfer terms at criticality in the whole range of the coupling constant
$b^{-d}$. The leading finite-size corrections to $d_2$ decay algebraically with
exponents equal to $-1$. At the infinite-size limit, it is found that $d_2$ is
of the form $d_2=c_db^d$ for small $b^d$ and $d_2=d-a_d/b^d$ for large $b^d$.
$P_c(s)$ is found to be scale-independent at all values of $b^{-d}$.
The large $s$ part of $P_c(s)$ obtained is shown to have an
$\exp (-A_ds^{\alpha})$ decay with $1 \le \alpha \le 2$. Finally, we determined
the disorder dependence of $\alpha$ in both the strong ($b^d \ll 1$) and the weak
($b^d \gg 1$) coupling regimes. At the limit of very strong disorder $b^d \to 0$,
we found that $\alpha \to 1$ and so we obtain the expected results of the
Poissonian decay predicted in Refs. \cite{SS93,AZ88}.

\begin{acknowledgement}
The author thanks the Spanish DGESIC for financial support through project
numbers BFM2003-03800 and FIS2004-03117.
\end{acknowledgement}

\end{document}